\documentstyle[12pt]{article}
\newcommand{\mysection}[1]{\setcounter{equation}{0}\section{#1}}

\newcommand{\nc}{\newcommand}
\nc{\beq}{\begin{equation}} \nc{\eeq}{\end{equation}}
\nc{\beqa}{\begin{eqnarray}} \nc{\eeqa}{\end{eqnarray}}
\nc{\lsim}{\begin{array}{c}\,\sim\vspace{-21pt}\\< \end{array}}
\nc{\gsim}{\begin{array}{c}\sim\vspace{-21pt}\\> \end{array}}

\begin{document}

\begin{titlepage}

\begin{center}
{\hbox to\hsize{hep-th/9507169 \hfill EFI-95-44}}
{\hbox to\hsize{\hfill Fermilab-Pub-95/258-T  }}
{\hbox to\hsize{July 1995, revised \hfill }}
\bigskip

\bigskip

\bigskip

\bigskip

{\Large \bf  Some Examples of Chiral  Moduli Spaces and }

\bigskip

{\Large \bf Dynamical Supersymmetry Breaking}

\bigskip

\bigskip

{\bf Erich Poppitz}$^{\bf a}$  and  {\bf Sandip P. Trivedi}$^{\bf b}$ \\

\bigskip

\bigskip

$^{\bf a}${\small \it Enrico Fermi Institute\\
 University of Chicago\\
 5640 S. Ellis Avenue\\
 Chicago, IL 60637, USA\\
{\rm email}: epoppitz@yukawa.uchicago.edu\\}

\smallskip

 \bigskip

$^{\bf b}${ \small \it Fermi National Accelerator Laboratory\\
  P.O.Box 500, Batavia\\
  IL 60510, USA\\
  {\rm email}: trivedi@fnth05.fnal.gov\\}

\vspace{2cm}

{\bf Abstract}
\end{center}

\begin{quotation}
We investigate the low-energy dynamics of $SU(N)$ gauge theories with
one antisymmetric tensor field,
$N - 4 + N_f$ antifundamentals and $N_f$ fundamentals, for $N_f \le 3$.
For $N_f = 3$ we construct the quantum moduli space, and for $N_f < 3$ we
find the exact quantum superpotentials.
We find two large classes of models with dynamical supersymmetry
breaking.
The odd $N$ theories break supersymmetry once appropriate mass terms are
added in the superpotential.
 The even $N$ theories break supersymmetry after gauging
an extra  chiral $U(1)$ symmetry.

\end{quotation}

\bigskip

\end{titlepage}

\renewcommand{\thepage}{\arabic{page}}
\setcounter{page}{1}

\baselineskip=18pt

\mysection{Introduction}

There are two motivations for studying non-perturbative supersymmetry (SUSY)
breaking. Firstly, it could explain why the electroweak
scale ($M_W$) is
so much smaller than the GUT or Planck scale ($M_{Pl}$). This
could happen if the
supersymmetry breaking scale is tied to the electroweak breaking
scale \cite{wittenDSB}.
The non-perturbative breaking would then relate $M_W/M_{Pl}$
to the logarithmic running of dimensionless coupling constants.
Supersymmetry would thus provide an explanation for both the fine tuning
and naturalness problems associated with the ratio $M_W/M_{Pl}$.
Secondly, since only chiral gauge theories can undergo dynamical
supersymmetry breaking, its
 study could shed some light on the
behaviour of non-perturbative chiral gauge theories - a subject
of interest from other points of view as well.

The past year has seen some spectacular progress in our understanding
of the non-perturbative behaviour of SUSY gauge theories (for a review
see \cite{seibergreview} and references therein).
 Most of it has been in   vector-like theories. In this paper we
extend these ideas to some chiral gauge theories as well,
uncovering in the process several examples of dynamical
supersymmetry breaking.
We will present some of our important results along with a few
details here. More results and details will follow in a subsequent paper.

Our general strategy is as follows. We will  restrict ourselves to theories
in which the scale of supersymmetry breaking is lower than the strong
coupling scale  ($\Lambda$)
of the gauge theory\footnote{In practice we
will achieve this by adding appropriately small
terms in the superpotential.}.
In such theories
the heavy degrees of freedom can be integrated out at the
scale $\Lambda$ and a supersymmetric  effective theory  can be
constructed in terms of the light fields. This effective theory can then
be used to study the breaking of supersymmetry.
If one is interested in explicitly calculating the
vacuum energy and the spectrum, knowledge of
both the K\"ahler potential and superpotential
are necessary.
However the K\"ahler potential is not needed in
detail if one only  wants to show that supersymmetry is broken.
For this purpose it is enough to ensure that the K\"ahler potential has no
singularities in terms of the light fields, i.e.  the moduli.
Supersymmetry breaking can then be established by analyzing the
superpotential.
Fortunately, a great deal can be said about the superpotential
non-perturbatively, while the K\"ahler potential in $N=1$ SUSY
gauge theories remains poorly understood.
Guided by these observations we first identify the correct moduli
fields in the low-energy effective lagrangian and argue that the K\"ahler
potential  does not contain any singularities (strictly speaking, we
will only
establish the absence of singularities for finite values of the moduli).
Then we turn our attention to the superpotential and
investigate the question of supersymmetry breaking. The strategy discussed
above is very similar to that of Intriligator, Seiberg and Shenker
\cite{ISS}.

In this paper, we consider some simple chiral
gauge theories \cite{ADS} - those with gauge group $SU(N)$ and one
antisymmetric tensor field, $N_f$ fields in the fundamental and
$N - 4 + N_f$  fields in the antifundamental
representations. While the analysis is quite different
for $N$  odd and even,  we find that in  both cases for $N_f = 3$
 these theories have a smooth moduli space
in terms of appropriately identified variables. The classical singularities
(present for example at the point where the gauge symmetry is restored) get
smoothed out quantum mechanically. Our results in this regard are analogous
to those obtained by Seiberg  for
SUSY QCD with $N_f = N_c$ \cite{seibergexact}.

Having established the quantum moduli spaces for these cases, we then add
various terms
 to the superpotential and study the behaviour of these theories.
In particular, by adding  mass terms we find the quantum superpotentials
for theories with $N_f \le3$.
In fact we find that for appropriate mass terms the $SU(2k+1)$ theories
break supersymmetry. This yields a large class of models which exhibit
supersymmetry breaking, some
of which (with $N_f=2$) are calculable.
For the $SU(2k)$ case we have not found any examples of SUSY breaking
in this manner. However by starting with
the $SU(2k+1)$ theories and breaking the gauge symmetry down
to $SU(2k)\times U(1)$ one arrives at a closely related set of theories.
Witten index considerations suggest they might break SUSY.
An investigation shows that with appropriate Yukawa terms this is indeed so,
thereby uncovering another class of models that dynamically break
supersymmetry.

This letter is organized as follows.
In sections 2, 3 and 4  we consider the $SU(2k+1)$, $SU(2k)$
and $SU(2k)\times U(1)$ theories respectively.
We end in  section 5 with conclusions and some comments.

\mysection{SU(2k+1)}

In this section we consider the nonperturbative low-energy dynamics of
models based on the gauge group $SU(2 k + 1)$, with an antisymmetric
tensor $A_{\alpha \beta}$, $2k - 3 + N_f$
antifundamentals $\bar{Q}^{\alpha}_i$,
 ($i = 1,...,2k - 3 + N_f$),
 and $N_f$ fundamentals $Q^a_\alpha$, ($a = 1,...N_f$). In this paper
we restrict  ourselves to $N_f \le 3$.
 In the absence of  any superpotential the classical
$SU(N)$ theory has a global
$ SU(N_f)_L \times SU(N - 4 + N_f)_R \times U(1)_Q \times
U(1)_{\bar{Q}} \times U(1)_A \times U(1)_R$ symmetry. The
charges of the fundamental  fields and the coefficients of the
anomalies (i.e. the ''charges" of the strong coupling scale
$\Lambda^{b_0}$,
with $b_0 = 2 N + 3 - N_f$ the first coefficient
of the beta function) of  the U(1)-symmetries are:
\begin{equation}
\label{charges}
\matrix{   &U(1)_Q& U(1)_{\bar{Q}}&U(1)_A &U(1)_R  \cr
Q            & 1   & 0     & 0    & 0  \cr
\bar{Q}    & 0   & 1       & 0    & 0  \cr
A             & 0   & 0           & 1  & 0  \cr
\Lambda^{b_0}&N_f   & N - 4 + N_f   & N - 2 & 6 - 2 N_f~. \cr}
\end{equation}

We start with the $N_f = 3$ case.
Subsequently, we will add mass terms and flow to theories
with a fewer number of flavors.
The classical  $N_f = 3$ theory has flat directions and hence infinitely many
inequivalent ground states. These flat directions
can be described
by the following gauge invariant operators\footnote{It can be shown that
all other invariants in this theory are products of the ones given
in (\ref{oddinvts}).}:
\beqa
\label{oddinvts}
M_i^a ~&=&~ \bar{Q}^\alpha_i~ Q_\alpha^a \nonumber \\
X_{ij}  ~&=&~A_{\alpha \beta} ~\bar{Q}^\alpha_i ~\bar{Q}^\beta_j \nonumber \\
Y^a ~&=&~ Q_{\alpha_{2 k + 1}}^a ~\epsilon^{\alpha_1 ...\alpha_{2 k + 1}}
{}~A_{\alpha_1 \alpha_2} ... A_{\alpha_{2 k - 1} \alpha_{2 k}}  \\
Z ~&=&~ \epsilon^{\alpha_1 ...\alpha_{2 k + 1} }~
A_{\alpha_1 \alpha_2} ... A_{\alpha_{2 k - 3} \alpha_{2 k -2 }}
{}~Q^a_{\alpha_{2 k - 1}}~ Q^b_{\alpha_{2 k}} ~ Q^c_{\alpha_{2 k + 1}}~
\epsilon_{a b c} \nonumber ~.
\eeqa
The inequivalent ground states correspond to different expectation values of
these moduli.   Not all the fields in (\ref{oddinvts}) are
independent. There is one constraint relating them, which follows from
the Bose symmetry of the superfields. It is
given by\footnote{This constraint is also needed to correctly
account for the total number of degrees of freedom.}:
 \beq
\label{oddconstraint}
Y \cdot M^2 \cdot X^{k - 1} ~=~ c ~Z~ {\rm Pf} X ~,
\eeq
where ${\rm Pf} X \equiv \epsilon^{i_1 ... i_{2 k}} X_{i_1 i_2} ...
X_{i_{2 k - 1} i_{2 k}} $, $Y \cdot M^2 \cdot X^{k - 1} \equiv
\epsilon_{a b c} Y^a M^b_{i_1} M^c_{i_2} \epsilon^{i_1 ... i_{2 k}}
 X_{i_3 i_4} ... X_{i_{2 k - 1} i_{2 k}}$ and $c = k/3$.

Symmetry arguments show that for $N_f=3 $
no superpotential can be generated dynamically.
Thus the vacuum degeneracy must
persist and the quantum theory must have a moduli space of ground states.
Considerations similar to those in supersymmetric QCD with
$N_f = N_c$ \cite{seibergexact} suggest that
the classical constraint (\ref{oddconstraint}) is modified by
non-perturbative
effects, and becomes
\beq
\label{oddquantumconstraint}
Y \cdot M^2 \cdot X^{k - 1} ~-~c~ Z ~
{\rm Pf} X ~=~\Lambda^{4 k + 2}~,
\eeq
with $\Lambda$ being the strong coupling scale of the theory.

As in the case of supersymmetric QCD this quantum
modification meets several non-trivial tests.
For example the fields
on the quantum-deformed moduli space saturate the 't Hooft conditions
for the unbroken global symmetries at various points of enhanced symmetry
(the maximal enhanced symmetry is
$SU(3)_L\times SP(2k)_R\times U(1)_R \times U(1)$).
Furthermore, as we show below, on integrating out one  of the quark flavors
the instanton generated superpotential for the
$N_f=2$ case is correctly reproduced. We regard these tests as fairly
persuasive and so will assume that the modified constraint
(\ref{oddquantumconstraint}) is correct.
This  constraint can be implemented by
adding a term in the superpotential of the form:
\beq
\label{wodd3}
W_{N_f = 3} ~=~ L~(~Y \cdot M^2 \cdot X^{k - 1} ~-~
c~ Z ~{\rm Pf} X ~-~\Lambda^{4 k + 2}~)~,
\eeq
with $L$ being a Lagrange multiplier.

The low-energy effective Lagrangian can then be described in terms of
the moduli fields (\ref{oddinvts}) subject to the constraint
(\ref{wodd3}).
The quantum modification to the constraint  (\ref{wodd3}) results
in smoothing out the singularities present classically at points of partially
enhanced gauge symmetry.
Therefore no fields other than the moduli become massless in any
finite region of the
quantum moduli space. Since singularities of the K\"ahler potential are
due to the appearance of extra massless states,
we are lead to conclude that the
K\"ahler potential in terms of
fields (\ref{oddinvts}) is not singular for any
finite values of the moduli.

One caveat needs
to be added to the discussion of the previous
paragraph. Strictly speaking, the points of partially
enhanced gauge symmetry are removed
from the quantum moduli space (\ref{oddquantumconstraint}) only for
finite moduli vevs. But these points can still be reached when some moduli
become infinite (while others go to zero,
 in a way consistent with the quantum-modified
 constraint)\footnote{We thank M. Dine and Y. Shirman for a
related discussion.}.
In this limit some subgroup $H$ of the gauge group is restored,
with both the $H$ gauge coupling and the $H$-breaking vevs tending to zero.
The massless, weakly coupled gauge bosons of the restored
gauge group descend into the low-energy theory, causing a singularity in the
 K\"ahler potential\footnote{This weak-coupling
singularity can be explicitly
seen in some $N=2$ models, where the K\"ahler potential is known
\cite{argyresdouglas}.}.
The correct
low-energy degrees of freedom are then given by the weakly coupled
quark and vector superfields. We will need to worry about these singularities
at infinity  in our discussion of supersymmetry breaking below.

Having understood the quantum moduli space and identified the appropriate
moduli fields we now turn to perturbing this theory by adding various terms
to the superpotential.
By adding a mass perturbation for the third flavor and integrating
out the heavy fields, we find the superpotential for the $N_f = 2$
theory :
\beq
\label{oddnf2superpotential}
W_{N_f = 2} ~=~ { \Lambda_{(2)}^{4 k + 3}
 \over \epsilon_{a c}~ Y^a ~M^c_{j_1}~
\epsilon^{j_1 ... j_{2 k - 1}} ~
 X_{j_2 j_3} ... X_{j_{2 k - 2} j_{2 k - 1}} }
\eeq
where we have absorbed a numerical coefficient in the definition of the
low-energy $\Lambda$ ($\Lambda_{(2)}^{4 k + 3} = m \Lambda^{4 k + 2} $).
The fields appearing in $W$ are the $N_f = 2$ analogues of
 the fields appearing
in (\ref{oddinvts}).
This superpotential has the simple physical interpretation
of being induced by a one instanton term in the gauge theory.

Let us now perturb the $N_f = 2$ superpotential (\ref{oddnf2superpotential})
by  adding  mass and Yukawa terms
\beq
\label{massperturbation}
\delta W = m_c^i M^c_i ~+~\lambda^{ij} X_{ij}.
\eeq
If SUSY is to remain unbroken, the
superpotential must be an extremum with respect to all the fields.
On extremising with respect to the mesons $M^c_i$ we find that :
\beq
\label{eqnmotnm}
m_c^i ~= ~ {\Lambda_2^{4 k + 3}
\over ( ~Y \cdot M \cdot X^{(2 k - 2)/3}~)^2}~ Y_c ~ \epsilon^{i j_2 ...
j_{2 k - 1}} ~X_{j_2 j_3} ... X_{j_{2 k - 2} j_{2 k - 1}} ~.
\eeq
But these equations cannot be satisfied for
a rank 2 mass matrix.
To see this consider starting with a diagonal mass matrix
(with the index $i$ in (\ref{massperturbation}) taking only two values,
$2k-2$ and $2k-1$ respectively). Then (\ref{eqnmotnm})
can be satisfied only if two contradictory conditions are met:
\beq
\label{yratio}
{ Y_1 \over Y_2} ~=~ 0, ~~{\rm and}~~~ { Y_2 \over Y_1} ~=~ 0~.
\eeq
Clearly this is impossible for any values of $Y_1$ and $Y_2$.
Since we have already argued that the K\"ahler potential has no
singularities for finite values of the moduli
\footnote{Although this was strictly shown for the theory with
$N_f = 3$ and without any mass terms in the superpotential, we expect
the conclusion to be true even after the mass terms are added.
After giving mass to some fields one expects some of the moduli fields
to get mass rather than extra massless states to appear. In fact, arguments
similar to the ones following eq.(\ref{eqnmotnm}) hold for the three-flavor
case as well.}, we conclude that
there is no SUSY preserving minimum in this region of moduli space. However,
to establish SUSY breaking we need to also rule out the possibility that
SUSY is restored when some of the moduli go to infinity. As was discussed
above, sometimes the K\"ahler potential may be singular along such runaway
directions. Thus, even though the superpotential cannot be extremized, the
vacuum energy may vanish and SUSY may be restored.
However, since these theories with
the superpotential (\ref{massperturbation}) have no
classical flat directions, the appearance of such runaway directions is
extremely improbable. We thus expect that these theories break supersymmetry.

We close this section with a few comments.

Firstly, for small $m \ll \Lambda$, $\lambda \ll 1$, the minimum
of the scalar potential of the $N_f = 2$ theory is expected to be in
the semiclassical region and the resulting models are therefore
calculable.
Readers primarily interested  in examples of SUSY breaking may note  that
these models can be understood simply without recourse to the
preceding discussion of moduli space etc. In this case the gauge symmetry
is completely broken and the
non-perturbative superpotential can be understood as simply arising from a
a single instanton effect.

Secondly, it has long been suggested \cite{ADS}, \cite{ADS2}, that
the $SU(2 k + 1 )$
theories with $2 k - 3$ antifundamental fields break supersymmetry.
These theories can be thought of as the $m \gg \Lambda$ limit of the
theories discussed above. In this limit we cannot strictly
speaking make any definite statements, nevertheless our results indicate that
supersymmetry breaking persists in this case as well\footnote{
Murayama\cite{Murayama} had suggested that the $SU(5)$ model in this class
can be analysed by adding an extra flavor with a mass term. Our analysis
above is very close in spirit to his. We have chosen to elaborate
on the $N_f=2$ case
since this yields calculable models.}.

\mysection{SU(2k)}

In this section we investigate the even-$N$
theories with antisymmetric tensors. We keep the discussion
brief since it closely parallels that in the previous section.
Once again we restrict ourselves to $N_f \le 3$.
We find a quantum moduli
space for $N_f = 3$, and  dynamically generated superpotentials
for $N_f < 3$.
{}From the point of view of constructing models for SUSY breaking the
results of this section will be primarily interesting as a stepping stone
for constructing the $SU(2k)\times  U(1)$ models discussed in the
next section.

The fundamental fields of the  $SU(2 k)$ theory are the
 antisymmetric
tensor $A_{\alpha \beta}$, $2k - 4 + N_f$
antifundamentals $\bar{Q}_i^\alpha$,
 ($i = 1,...,2k - 4 + N_f$),
 and $N_f$ fundamentals $Q^a_\alpha$, ($a = 1,...,N_f$).
The classical moduli space for $N_f = 3$
 is described by the following gauge invariant
fields\footnote{As in the odd case,
one can show that all other
invariants are products of (\ref{eveninvts}).}:
\beqa
\label{eveninvts}
M_i^a ~&=&~ \bar{Q}^\alpha_i~ Q_\alpha^a \nonumber \\
X_{ij}  ~&=&~A_{\alpha \beta} ~\bar{Q}^\alpha_i ~\bar{Q}^\beta_j \nonumber \\
Y_a ~&=&~\epsilon_{a b c} Q^b_{\alpha_1} Q^c_{\alpha_2}
 ~\epsilon^{\alpha_1 ...\alpha_{2 k}}
{}~A_{\alpha_3 \alpha_4} ... A_{\alpha_{2 k - 1} \alpha_{2 k}}  \\
{\rm Pf} A ~&=&~ \epsilon^{\alpha_1 ...\alpha_{2 k } }~
A_{\alpha_1 \alpha_2} ... A_{\alpha_{2 k - 1} \alpha_{2 k}}\nonumber ~.
\eeqa
These invariants are subject to a constraint, which is modified by
nonperturbative effects and becomes:
\beq
\label{evenquantumconstraint}
X^{k - 1} \cdot M \cdot Y ~- ~b ~
M^3 \cdot X^{k-2} ~{\rm Pf} A~=~\Lambda^{4 k}~~.
\eeq
Here
$M^3 \cdot X^{k - 2} \equiv \epsilon_{a b c} M^a_{i_1} M^b_{i_2} M^c_{i_3}
 \epsilon^{i_1 ... i_{2 k -1}} X_{i_4 i_5} ... X_{i_{2 k - 2} i_{2 k - 1}}$,
$X^{k - 1} \cdot  M \cdot Y \equiv Y_a M^a_{i_1} X_{i_2 i_3}$ ... $
X_{i_{2 k - 2} i_{2 k - 1}} \epsilon^{i_1 ... i_{2 k - 1}}$ and
$b = (k - 1)/(3 k)$. 't Hooft's
anomaly
matching conditions are saturated
by the moduli fields subject to the constraint (\ref{evenquantumconstraint})
(the maximal enhanced symmetry in this case
 is $SU(3)_V\times SP(2 k - 4)\times
 U(1)_R$).

The dynamical superpotentials for $N_f < 3$
\footnote{The moduli fields appearing in the superpotentials for $N_f < 3$
are simply the restrictions of the $N_f = 3$ invariants
 (\ref{eveninvts})to a
smaller number of
flavors.}
can be found by integrating out
the extra flavors. For $N_f = 2$ we find the instanton induced superpotential
\beq
\label{weven2}
W_{N_f = 2} ~=~ {\Lambda_{(2)}^{4 k + 1} \over
Y {\rm Pf} X -~3 b ~ \epsilon_{a c}
 M^a_{i_1} M^c_{i_2} \epsilon^{i_1 .. i_{2 k - 2} }
 X_{i_3 i_4} ... X_{i_{2 k - 3} i_{2 k - 2}} {\rm Pf} A }~.
\eeq
The singularity at $Y ~{\rm Pf} X = M^2 \cdot X^{k - 2} {\rm Pf} A$
 reflects the existence of
points on the moduli space  with an  unbroken $SU(2)$ gauge symmetry. For
$N_f = 1$ the superpotential is due to gaugino condensation in the unbroken
$SU(2)$ gauge group:
\beq
\label{weven1}
W_{N_f = 1} ~=~{ ~\Lambda_{(1)}^{2 k + 1} \over
\left[~ M_{i_1} X_{i_2 i_3} ... X_{i_{2 k - 4} i_{2 k - 3}}
\epsilon^{i_1 ...i_{2 k - 3}} {\rm Pf} A ~  \right]^{1/2}~}~.
\eeq
Finally, the $N_f = 0$ superpotential is induced by SP(4) gaugino
condensation \cite{ADS}:
\beq
\label{weven0}
W_{N_f = 0} ~=~ {~\Lambda_{(0)}^{(4 k + 3)/3} \over
\left[~ {\rm Pf} A ~{\rm Pf} X ~\right]^{1/3}~}.
\eeq

We have investigated these theories by adding
various terms to the superpotentials, (\ref{weven2}), (\ref{weven1}) and
(\ref{weven0}), but have not found any examples of SUSY breaking.
For example on adding a perturbation
$\delta W = {\rm Pf} A + \lambda^{ij} X_{ij}$ to (\ref{weven0}) we can
see that the $N_f = 0$
theory has a supersymmetric ground state.

\mysection{SU(2k)$\times$U(1)}

In this section we investigate the behaviour of theories with gauge group
$SU(2k) \times U(1)$ and find a large class of models that do break SUSY.
These theories can be obtained by starting with the
$SU(2k+1)$ theories of section 2,
and breaking the gauge symmetry down to $SU(2k) \times U(1)$.
The symmetry breaking can be accomplished,
for example,  by adding an  additional heavy field in the adjoint of
$SU(2k+1)$.
Since the $SU(2k+1)$ theory is expected to break SUSY
it has a zero Witten index. This would suggest that the resulting $SU(2k)
\times U(1)$ theory has a vanishing Witten index too, thereby making
it a natural candidate for SUSY breaking. As we show below SUSY
breaking does indeed occur in these models.

Our starting point will be the $SU(2k+1)$ models of section
2 with $N_f = 0$.
The $SU(2k)\times U(1)$ theory resulting from
symmetry breaking is then  given by the $SU(2k)$ theory with $N_f = 1$
of section 3 with additional $2 k - 3$ $SU(2k)$ singlets, $S_i$,
($i = 1,...,2 k - 3$). The U(1) charges of the fields are
$\bar{Q} \sim -1$, $S \sim 2 k$, $A \sim 2$, $Q \sim 1 - 2 k$.

We will only consider theories where the
$U(1)$ gauge coupling is weak at the scale at which the
$SU(2k)$ coupling gets strong\footnote{Since the U(1) coupling is
irrelevant in the infrared it only gets weaker at lower energies.}.
In this case the low energy lagrangian can be simply constructed in two
steps. First one can neglect the $U(1)$ interaction
and construct the effective lagrangian of the $SU(2k)$ theory. Then
one can gauge the $U(1)$ symmetry in this lagrangian, integrate out the
resulting heavy particles and construct the final low energy lagrangian.
Since the first step was already carried out in section 3 (the
additional $SU(2k)$ singlets clearly do not pose any problems)
we can directly turn to gauging the $U(1)$ symmetry in the lagrangian
containing the $SU(2k)$ moduli fields, (\ref{eveninvts}) and the
singlets, $S_i$.

If the $U(1)$ symmetry  is broken the relevant degrees of freedom
in the low energy lagrangian are the
$U(1)$ invariants built out  of the $SU(4)$-invariant moduli.
The $SU(2k)$
invariant moduli fields have  charges ${\rm Pf} A \sim 2 k$,
 $M_i \sim - 2 k$, $X_{ij} \sim 0$,
 and $S_i \sim 2 k$ under the $U(1)$ symmetry.
Out of them we can build
three types of  $SU(2k)\times U(1)$ invariants
\beq
\label{u1invts}
A_i ~=~  M_i~{\rm Pf} A , ~~ B_{ij} ~=~ S_i M_j ,  ~~ {\rm and } ~~ X_{ij} ~~.
\eeq
These fields  are not all independent but
obey the constraints
\footnote{ These constraints are  not all independent.
However, adding a redundant set of constraints only amounts
to redefining the Lagrange multipliers for the independent
constraints and does not alter any of the conclusions.}
\beq
\label{u1constraint}
B_{ij} A_k - B_{ik} A_j = 0.
\eeq
Unlike the cases encountered previously, these constraints are
not modified quantum - mechanically. This is expected due to the
non-asymptotically free nature of the $U(1)$ gauge
interaction and  can also be
seen explicitly by symmetry considerations.
The moduli fields in the final low energy theory are thus given by
$X_{ij}$, $A_i$ and $B_{ij}$ subject
to (\ref{u1constraint}).
The K\"ahler potential
in terms of these fields has  singularities which occur
on the submanifold where the $U(1)$ symmetry is restored. There extra massless
particles (e.g. the $U(1)$ gauge boson) descend into the
low energy theory resulting in the singular K\"ahler potential. In
our analysis of SUSY breaking we will have to consider this submanifold
separately.

The dynamical superpotential (\ref{weven1}) can be written in terms of the
U(1) invariant fields as:
 \beq
\label{even1}
W_{dyn} ~=~ { \Lambda_{(1)}^{2 k + 1} \over
\left[~ A_{i_1} X_{i_2 i_3} ... X_{i_{2 k - 4} i_{2 k - 3}}
\epsilon^{i_1 ...i_{2 k - 3}}~ \right]^{1/2}~}~.
\eeq
Let us add to this the Yukawa couplings
\beq
\label{suptree}
W_{tree} = \gamma^{ij} X_{ij} + \lambda^{ij} B_{ij},
\eeq
and implement the constraints (\ref{u1constraint})
via a Lagrange multiplier,
$$W_{constr} ~=~ L^{i}_{l_1 ... l_{2 k - 5}}
{}~\epsilon^{l_1 ... l_{2 k - 3}} ~B_{i l_{2 k - 4} } A_{l_{2 k - 3}}~. $$

We are now ready to show that the $SU(2k)\times U(1)$
  model breaks supersymmetry.
For simplicity, we first take $k = 3$. In this case
 $W_{constr} = L^i_j \epsilon^{j k l} B_{i k} A_l$ and
the equations of motion
for $B_{ij}$ are
$ L^i_k  \epsilon^{k l j}  A_l =  \lambda^{ij} ~.$
Solving the $i = 2, 3$ equations  for  $L^2_k$ and $L^3_k$,
and substituting back into the $i = 1$ equation we find three conditions;
$ \lambda^{l 1} + \lambda^{l 2}  (A_2/ A_1) +  \lambda^{l 3}
(A_3/A_1) =  0 $, for $l = 1,2,3$. Clearly these cannot  be satisfied
  when $\lambda^{ij}$ has rank three (to
see this consider going to a basis where
$\lambda^{ij}$ is diagonal).
 This argument can now be easily generalized
for $k > 3$. Solving the equations of motion for the Lagrange
multipliers one obtains a similar consistency condition
 $ \lambda^{i 1} + \sum\limits_{j = 2}^{2 k - 3}
\lambda^{i j} (A_j / A_1) = 0$, $i = 1,..,2 k - 3$,
 which again  cannot be  satisfied by a
non - degenerate Yukawa matrix. Thus,  in all these models
 with nondegenerate  Yukawa couplings we expect  that
 SUSY is broken. We should add though, that, as in the $SU(2k+1)$ case
we cannot strictly rule out the possibility of runaway directions. Such
directions might arise since the K\"ahler potential can become singular
when some of the moduli vevs go to infinity. The vacuum energy then
may go to zero, even though the superpotential is not extremized. However,
as in the $SU(2k+1)$ case, the $SU(2k)\times U(1)$ theories with
superpotential (\ref{suptree}) have no classical flat directions and we do not
expect such runaway directions to be induced quantum-mechanically. We
thus expect that supersymmetry is broken.

As mentioned earlier, we need to consider the submanifold
on which the $U(1)$ symmetry is restored separately (on this submanifold
the vevs of all $U(1)$-charged $SU(2k)$ moduli,
$M_i$, $S_i$ and ${\rm Pf} A$,
go to zero).
The correct degrees of freedom
around any point in this submanifold are the $SU(2k)$ moduli fields
 and the $U(1)$ gauge field. By   varying the full superpotential,
which is the sum of (\ref{even1}) and (\ref{suptree}),
 with respect to the $SU(2k)$ moduli it is easy to see, that
there is no way in which
$S_i$,$ M_i$ and  ${\rm Pf} A$ can tend to zero
while preserving supersymmetry\footnote{This is true even if one allows
for runaway directions.}. Thus supersymmetry cannot be restored on this
submanifold and these theories do indeed break SUSY.

The case $k=2$ corresponds to the simplest model in this class.
It is worth discussing in some more detail.
The theory has a gauge group $SU(4) \times U(1)$, and only three
$SU(4)$ moduli fields, denoted by $M$, ${\rm Pf}A $ and $S$
respectively\footnote{In this case the field $X_{ij}$ (\ref{eveninvts})
 is absent.}. The full superpotential
(corresponding to a sum of the terms (\ref{even1}) and (\ref{suptree})
  ) is given by:
\beq
\label{W11}
W ~=~ {\Lambda_{(1)}^5 \over \sqrt{M ~~ {\rm Pf}~A} }~ +~\lambda S ~M~.
\eeq
The $U(1)$ invariants correspond precisely to the two combinations
$M ~{\rm Pf}A$ and $S ~M$ and  no constraints are needed in this
case. Extremising the superpotential with
respect to these fields clearly shows that SUSY is broken \footnote{We
do need to consider the points of restored $U(1)$ symmetry separately but
as above they do not change the conclusion.}.
This model is among the simplest examples of SUSY breaking we know of.
The superpotential
(\ref{W11})
 preserves an R-symmetry. On adding
another term, $M ~{\rm Pf} A$, one finds that
SUSY breaking persists even though the R-symmetry is now broken.
This is another example of supersymmetry breaking
without R-symmetry \cite{nelsonandseiberg}.

We end this section by commenting on  the
 importance of  correctly incorporating  constraints
(for example (\ref{u1constraint})) into the
analysis  when testing for SUSY breaking, especially with respect to
runaway directions. As a toy model,
 consider a theory with
the nonsingular K\"ahler potential $K = X^* X + Y^* Y$ and superpotential
$W = X + L( X Y - 1 )$.  If we first solve the constraint for $Y$, $Y = 1/X$,
and then minimize the superpotential with respect to
$X$, we would conclude that
the theory breaks supersymmetry since  $d W/d X = 1$. However,
 by solving the
constraint we introduced a singularity in the K\"ahler metric, $K_{X X^*}
\sim 1/( X^* X)^2$ at $X \rightarrow 0$, or $Y \rightarrow \infty$.
 Therefore
the inverse K\"ahler metric has a zero eigenvalue and the model has a
 runaway direction.  Had we kept the
constraint, this behaviour would follow solely
from the superpotential.

\mysection{Conclusion}

In this letter we studied the low-energy dynamics of chiral
$SU(N)$ gauge theories with one antisymmetric tensor and
$N - 4 + N_f$ antifundamental and $N_f$ fundamental fields.
We found the quantum moduli spaces and
exact superpotentials for the models with $N_f \le 3$.
We also found two large classes of models that broke
supersymmetry dynamically.
For the odd-N models this breaking occurred when suitable
mass terms were added to the superpotential. For the even-N models
the supersymmetry  breaking occurred after gauging  an additional
chiral $U(1)$ symmetry.
These results suggest that, perhaps, the set of theories
which undergo dynamical
supersymmetry breaking is quite large  and might even be
a fairly large subclass of all chiral SUSY gauge theories.

Clearly, much more needs to be done to further these investigations.
In the short run it would be interesting to extend this analysis
to a larger number of  flavors,   $N_f>3$, hopefully in the recently
proposed framework of duality
\cite{seiberg.witten}, \cite{seiberg},  \cite{berkooz}.
{}From a phenomenological point of view it would be
interesting to incorporate these theories into visible sector SUSY
extensions of the standard model \cite{DNold}.
In the longer run one would like to understand better the essential
ingredients required for supersymmetry breaking and attempt a more
systematic construction of the possibly large class of theories
that exhibit this phenomenon.

\bigskip

While completing this paper, we became aware of the recent preprint
\cite{DNSN}
where some of our results were obtained.

\section{Acknowledgements}

We are grateful to  D. Kutasov, J. Lykken and
L. Randall  for
useful discussions. E.P. acknowledges support by a Robert R. McCormick
Fellowship and DOE contract DF-FC02-94ER40818. S.T. acknowledges the
support of D0E contract DE-AC02-76CH03000.

\end{document}